# Large language models in climate and sustainability policy: limits and opportunities

Larosa, F.[1,*], Hoyas, S.[2], Conejero, J.A.[2], Garcia-Martinez, J.[3], Fuso Nerini, F.[4], Vinuesa, R.[1]

[1] FLOW, Engineering Mechanics, KTH Royal Institute of Technology, Stockholm, Sweden
[2] Instituto Universitario de Matemática Pura y Aplicada, Universitat Politècnica de València, Valencia, Spain
[3] Departamento de Química Inorgánica, Universidad de Alicante, Alicante, Spain
[4] Division of Energy Systems, School of Industrial Engineering and Management, KTH Royal Institute of Technology, Stockholm, Sweden
* Corresponding author: larosa@kth.se

**Abstract**
As multiple crises threaten the sustainability of our societies and pose at risk the planetary boundaries, complex challenges require timely, updated, and usable information. Natural-language processing (NLP) tools enhance and expand data collection and processing and knowledge utilization capabilities to support the definition of an inclusive, sustainable future. In this work, we apply different NLP techniques, tools and approaches to climate and sustainability documents to derive policy-relevant and actionable measures. We focus on general and domain-specific large language models (LLMs) using a combination of static and prompt-based methods. We find that the use of LLMs is successful at processing, classifying and summarizing heterogeneous text-based data. However, we also encounter challenges related to human intervention across different workflow stages and knowledge utilization for policy processes. Our work presents a critical but empirically grounded application of LLMs to complex policy problems and suggests avenues to further expand Artificial Intelligence-powered computational social sciences.

**1. Introduction**

The intertwined nature of multiple global crises is threatening the stability of our planet and is posing challenges to our societies. Direct and cascading climate impacts are provoking biodiversity collapse and ecosystem destruction, disinformation is posing democracy under attack and military and economic wars are redesigning the geopolitical landscape[1]. In the face of these issues, digital technologies, if properly designed and used, could support sustainability transitions (STs) by increasing the efficiency of the production and consumption modes[2], democratizing knowledge through access to information[3] and by launching new scientific breakthroughs[4]. The Internet of Things (IoT) monitor in quasi real-time the state of our connected world, while machine learning (ML) and artificial intelligence (AI) process and elaborate the wealth of information produced to generate new insights and knowledge.

In the sustainability realm, environmental, economic and social dimensions act simultaneously on anthropogenic and natural ecosystems. The interaction across different domains can be positive or negative, leading to synergies and trade-offs across different time and spatial scales. The study of these linkages has recently gained popularity: researchers have engaged in global[5], regional[6] and sectorial[7] analyses to turn interactions into actions. These efforts are possible also thanks to the 17 Sustainable Development Goals (SDGs), launched by the United Nations (UN) in 2015 and operationalized through 149 targets and 281 quantitative indicators. The SDGs offer a measurable and traceable universal roadmap for the near future (up to 2030) and detail what is at the heart of sustainable development. Since their establishment, the SDGs have spurred funding, legislation initiatives and policy prescriptions worldwide generating text-based inputs heterogeneous per nature, scope and aim. The analysis of these documents reveals different degrees of ambitions[8], critical gaps[9] and political priorities[10]. However, the elicitation of policy-relevant knowledge is often left to expert groups, extensive manual discourse and content analyses with hard-to-replicate routines[11]. Oftentime, findings are hard to elucidate as complex trade-offs and synergies are involved. To overcome selection biases and to boost open science practices, the use AI-based approaches – mainly natural-language processing (NLP) – has become prominent[12]. The widespread penetration of large language models (LLMs) supports this trend and opens the door to hybrid methods which integrate human and machine-based insights. As more studies flourish in this space, so do critiques[13,14] The strongest fronts call for more transparent, open and explainable models[15,16], but there are growing concerns about AI's environmental footprint and costs[17]. Beyond the sustainability and ethical considerations of the use of LLMs (and generative AI in general), the usability of AI outputs for public policy and policy agenda setting remains loosely.

In this work, we tackle this question by providing an empirical and critical overview of the use of LLMs for climate and sustainability policy purposes. We provide evidence of the *process* behind the transdisciplinary research around these topics and document where barriers exist in the deployment of NLP and LLM findings for policy. We do so by presenting the workflow of two parallel large-scale analyses which use different text-based corpora. The first includes political documents called Nationally Determined Contributions (NDCs) redacted to commit domestic efforts to climate adaptation and mitigation. The second corpus, which covers the entire spectrum of sustainability challenges, aims at assessing how different document types identify synergies and trade-offs across the 17 SDGs. Rather than presenting the sequential steps of both analyses, our work looks at the procedural aspects of both analyses, critically assessing whether the application of certain NLP and LLM are more suitable than others to elicit policy-relevant insights.

To answer this question, we take a step-wise approach. First, we discuss recent advancements in NLP and LLM application to sustainability challenges, complementing published literature on this matter with a focus on usability. Second, we present the trial-and-error process behind both studies and we group limits and research needs according to the different workflow stages. Finally, we propose three research avenues for the future to move from usefulness to usability of AI-generated insights in the climate and sustainability domain.

## 2. From content analysis to computational linguistic in policy

In his seminal "Syntactic Structures" (1957), Noam Chomsky expresses the aim of his scientific inquiry as the concern to solve "the problem of determining the fundamental underlying properties of successful grammars." His theories were pivotal in understanding how machines could represent, translate, and manipulate language, hence paving the road for text parsing and language processing. Keyword-based and pattern-matching early chatbots in the 1960s further expanded the field of computational linguistics and supported the understanding that language is designed as a set of universal predefined rules which can be mimicked and replicated. The subsequent evolution of the computational linguistic and NLP field benefitted from knowledge representation and frame-based semantic networks, which encapsulate how language is used within specific boundaries. Logic-based programming languages introduced the idea that language and communication work according to rules and predicate logic. It was only in the 1990s that statistical methods gained prominence thanks to the availability of large text corpora from which language patterns and recognition could be drawn. Frequency methods ( including bag-of-words methods and Term Frequency-Inverse Document Frequency[18]) revealed important features of unordered collections of texts and words, exposing how specific terms were used in a context. While often considered superseded, these methods remain of high relevance, especially in contexts where interpretability can win over complexity. In the case of information-extraction tasks, for example, frequency-based methods support the understanding of how certain stakeholders tackle punctual topics.

Over the last decade, the integration of frequency-based methods with deep learning approaches has revolutionized the field of computational linguistics and NLP. The introduction of the attention mechanisms[18] launched a new architecture - the transformer – which allows models to focus on specific sections of the input to generate outputs. Transformers understand logical intra-sequence relationships between different parts and produce outputs based on those. Relationships can be statistically inferred based on previous training – on a predefined large corpus on a predefined large corpus of diverse texts to learn from. The more specific the training set on top of the general vocabulary, the more domain and task-specific the output. The development and deployment of pre-trained large language models (PLMs e.g., BERT[19]), were instrumental to achieving higher performance and versatility in diverse tasks with minimal fine-tuning. At the heart, these large language models drive generative AI: the creation of coherent and context-relevant content thanks to the availability of vast datasets underneath that provide various ranges of language structures and styles. The scalability of transformer-based architectures allows the creation of adaptable applications which can be flexibly used for academic, policy and commercial uses.

The use of computational linguistic methods in climate- and sustainability-policy studies departs from the observation that language can signal change and spur action, especially in the political context. Language is also at the heart of the relationship between "the evidence system" (i.e., academic and research production) and "the policy system" (i.e., the government arena and the public policy design and discussion)[20]. Over the past three decades, the study of language usage has widespread in the climate and sustainability arenas evolving in a variety of methods and conceptual approaches[21] starting with content and discourse analysis. Content analysis is a systematic approach which aims at analyzing textual features within a coding frame[22]. Content analysis can be qualitative if the core question refers to specific meanings or quantitative whether the interests are grounded in statistical measurements (such as numerosity or frequency). Researchers have been using content analysis alone or in combination with other methods to assess climate beliefs[23], investigate climate-related online communication[24] and explore the evolution of ideas in climate policy[25]. Discourse analysis studies language within specific socio-cultural structures[21]. It is a systematic method concerned with how meaning is formed, hence aiming at describing and understanding the process as shaped by contextual factors. Concepts are related to others and their relationships are not static, but rather open and contingent. Discourse analysis can often be linked to an understanding of how communication and language used affect the surrounding world (critical discourse analysis)[26]. While content analysis simplifies the textual material by reduction, discourse analysis frames meaning within hegemonic categories, meant as dominant ideas of classes of thought.

Both content and discourse analyses are grounded within the social science and linguistic practice. Their limitations (i.e., low replicability, extensive manual work, skewed perspectives and biased coding routines) motivated the shift towards computational linguistic methods, including topic modeling, NLP and – recently – LLMs. Topic modeling has been widely deployed to survey big literature on climate change research[27], to assess climate polarization online[28] and to understand how different countries tackle the climate crisis[29]. NLP methods, including frequency-based approaches, have been deployed to analyse climate risks as reported by companies[30] and to compare if climate discourses match trends observed over other political debates[31].

## 3. Tackling the climate and sustainability policy space

In this work, we explore whether different NLP techniques – including the use of LLMs – can support policy-makers with relevant and usable insights in sustainability and climate policies. We test text-centered AI routines over two corpora. The first focuses on climate policy, while the second has a broader sustainability breath. In the first work, our aim is to provide actionable insights to support the new submission cycle of the NDCs: political documents which countries submit every five years to the United Nations Framework Convention on Climate Change (UNFCCC) with details about their short and medium-run plans to reduce greenhouse gas emissions in the atmosphere. As these are climate action plans, their spillovers with other sustainability dimensions determine new winners and losers, opportunities for transformative change and new potential inequalities. The goal of the second application is to provide a workable AI-based routine to detect synergies and trade-offs among the 17 SDGs. As presented in section 2, these links are usually detected using human expertise and long coding procedures. We aim at offering a fully replicable alternative to achieve two major objectives: i) the mapping of individual texts to relevant SDGs; and ii) the assessment of positive (synergies) and negative (trade-offs) interlinkages between them. This routine leaves room for practitioners and policy-makers to design tailored policies, but also enables researchers to explore new questions and unresolved issues.

### 3.1. Aligning climate and sustainable development actions

The Paris Agreement – signed by 196 parties in 2015 – imposes countries "to limit global average temperature increases to as close to 1.5 degrees Celsius as possible". In its Article 4, the treaty requires that each party prepares, explains and shares how it intends to respect this commitment. NDCs meet that request: they are political documents, submitted and updated every five years, containing pledges to pursue climate mitigation (i.e., reducing emissions) and put in place adaptation actions. Individual submissions define the degree of ambition of each party. As a collective corpus, instead, the NDCs determine whether the world is on track to achieve the long-term goals of the Paris Agreement. As pathways to change are by definition developed overtime, the analysis of submissions becomes a powerful tool to assess diverging or converging behaviors with respect to climate policy. Being political documents, the NDCs have been studied as coordination mechanism[32] and governance instruments[33]. As they explore how different sectors, groups and areas of an individual party will have to behave with respect to renewed climate ambition plans, the NDCs also reveal other important political priorities. For this reason, researchers have engaged in the identification of areas of misalignment between the climate and sustainable development agenda. When present, gaps between the two create significant hurdles which slow or halt mitigation and adaptation efforts or harm other socio-economic and environmental dimensions. As NDCs change overtime per content, style and ambition, replicable analysis protocols become crucial to assess comparability and to ensure approaches do not alter the results.

We use the NDCs as comprehensive climate action documents and we assign the content of each individual text to all pertinent and relevant SDGs. We do this to assess which sustainable development dimensions are most covered and which are instead significantly neglected. We further reveal the nature of interactions between the climate and sustainable development agendas identifying whether links between the two act as synergies (positive) or as trade-offs (negative). Our findings aim at informing the international community as a whole, but also both national and international development agencies. By revealing critical disconnection points, our findings suggest new targeted areas for convergence in international finance mechanisms. As high-income countries committed to mobilise 100 billion USD per year up to 2020 and reinforced this baseline up to 2025 (in COP28 in Dubai) in order to support the low-carbon transition in lower-income countries, a comprehensive understanding of synergies to support both climate and sustainable development agendas are key to prioritize those sectors and areas of intervention with the highest potential impact. Our approach to the NDCs starts with the data collection. Rather than surveying experts to check alignment points with the SDGs as in Fuso Nerini *et al.*[34], we focus on the first two submission rounds of the NDCs to zoom into the content of declared pledges of change. The focus on political documents has a forward-looking purpose: we move beyond assessing positive or negative interlinkages between the climate and development agenda and we present what the near-term future might look like if pledges were fully met. By doing so, we proactively signal where critical misconnections are and we inform the new wave of NDC submissions.

The downloaded each NDC in PDF format from the official UNFCCC NDC registry (https://unfccc.int/NDCREG). Prior to text processing, we developed a data analysis strategy identifying two parallel workflows. The first entails the deployment of two downstream task models to achieve classification and sentiment analysis using ClimateBERT, a climate-domain-adapted LLM[35]. ClimateBERT uses DistilROBERTA as a starting point and it builds on additional training using corporate disclosures, climate-related paper abstracts and private companies' reports. Downstream models include – among others – the classification of climate-related sentiments in climate-focused paragraphs according to three categories: 0 if the paragraph contains negative text with respect to climate adaptation and mitigation; 1 if it is neutral and 2 if there is an explicitly positive recall to actions which may be positive to climate adaptation and mitigation. We were interested in detecting paragraphs tackling one or more specific SDGs and to derive whether these were expressing positive or negative viewpoints on climate adaptation and mitigation. The second strategy entails the deployment of a general LLM, in our specific case Gemini 1.0 by Google Research[36], and the design of the most adequate prompt to mirror the research objectives explained in the first strategy. Gemini 1.0 pro was used through the Google AI Studio Application Program Interface (API) and the library *generativeai*. The high degree of interoperability leads to the possibility of changing two lines of code to upgrade to a new Gemini version. The process can be useful due to the very fast development of this field.

The first strategy determined the length of the textual unit of interest: as climateBERT-sentiment is trained on paragraphs, we splitted the full-length NDCs into meaningful paragraph-long chunks. We deployed a PDF-dedicated Python package (PyMuPDF) applying a set of rules to clean the text: the exclusion of blocks with <2 words/numbers and with >50% of numerical characters (out of the total); the exclusion of captions by removing blocks starting with the expressions ""figure|figura|fig|table|tableau|tabla|chapter|chapitre|capítulo|page|pagina|página" followed by a number + string. The list could be made more comprehensive (for example, in the Andorra report, they use "Gráfico" instead of "Figura" and the elimination of short sentences (<25 words) repeated more than five times. As we notice, paragraphs may be broken into lines, reducing our ability to extract meaningful information. To overcome this issue, we complement our text cleaning and paragraph split routine with a sentence segmentation algorithm that is able to combine different sentences using pre-trained transformer models. We deployed SpaCy (https://spacy.io/usage/linguistic-features#retokenization), which builds on syntactic dependency parsing: broken sentences are detected through a dependency tree built with tokens. The difference in distribution with and without SpaCy is very small (Figure 1a), but shorter paragraphs are joined with others, creating a more coherent flow.

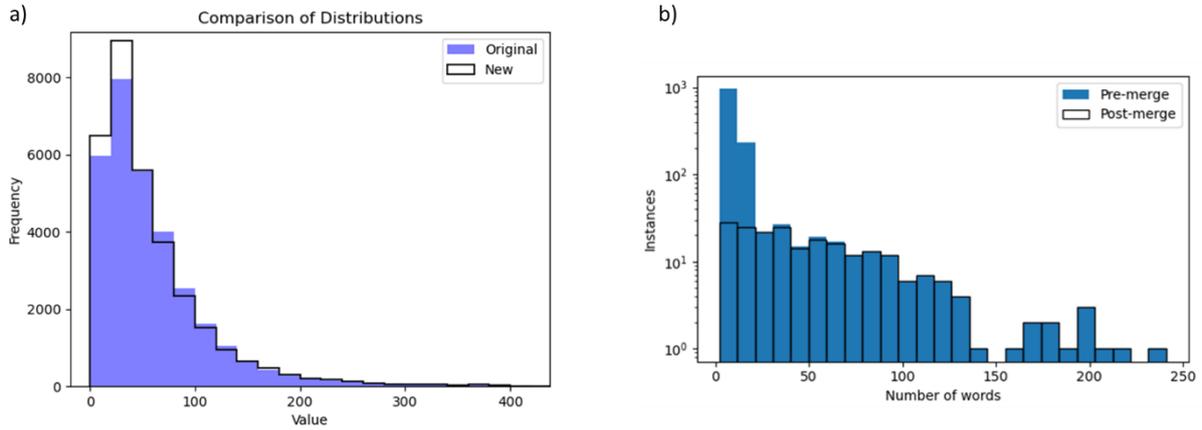

Figure 1. Comparison of "old" (pre-merged, without adding SpaCy routine) and "new" (post-merged) distributions. a) Paragraphs with a limited number of words are more frequent, but after merging the paragraphs become more comprehensive and move beyond the individual sentence length; b) An example: Andorra's NDC. Pre-merge, the NDC had 1358 paragraphs; after the merge, Andorra contains 221 paragraphs, 58 words long on average.

Once we obtained the clean and workable text, we proceeded to the SDG detection for every paragraph. We computed semantic similarity between every paragraph and the SDGs as derived from the UN SDG Fast Facts (https://www.un.org/sustainabledevelopment/sdg-fast-facts/). We used this description instead of any other alternative because it contains a more policy-friendly style and departs from pure academic language. We assigned to each paragraph the SDG with maximum similarity score and we then passed the so-obtained paragraphs into ClimateBERT-sentiment (Figure 2).

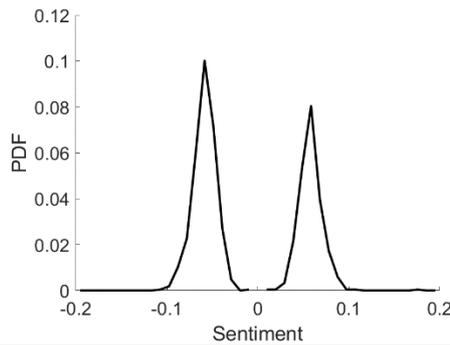
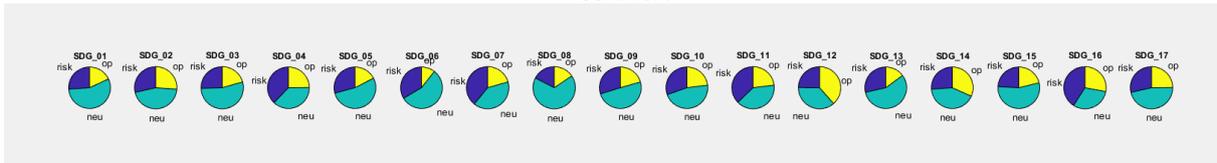

Figure 2. The sentiment and SDG allocation of the first strategy. SDGs are detected using semantic similarity between NDCs' paragraphs and SDG Fast Facts downloaded from the UN website. Sentiment (between 0 and 2) is then transformed into a probability distribution of the expected values of each category.

For our second strategy, we used Gemini 1.0. Our choice was motivated by a heuristic process grounded in linguistics and meaning. We developed three versions of the same data, and we applied it to two performance-equivalent models (GPT3.5 and Gemini 1.0). Each version carries a slight but meaningful-relevant modification (Table 1). Initially, we asked the models to reveal the top three SDGs, only focusing on three diverse classes: quantity (*dominance*), value (*relevance*), and a mix of the two (*prominence*). Dominance refers to the appearance of concepts and words in a given text. We consider it as more context-agnostic in this setting. Relevance is context-relevant and includes value judgments and value assessments of individual concepts. Even when less abundant in the text, some words can leave a mark on the reader because they are at the heart of a whole message. Prominence, in this contest, is a mixed area where words and concepts are both heavily cited and relevant.

| Table 1. Alternative prompts | |
|---|---|
| Prompt #1: quantity-oriented | "Assign the following text to the top three SDGs based on their **dominance**" |
| Prompt #2: value-oriented | "Assign the following text to the top three SDGs per **relevance**" |
| Prompt #3: quali-quantity oriented | "Assign the following text to the top three SDGs based on their **prominence**" |

We run each prompt three times on a sample of paragraphs as derived by the NDCs in random order. Our aim was twofold. First, we wanted to check whether the models capture the marginal but meaning-relevant modifications providing different results. Second, we aimed to assess the robustness of the models when identifying the relevant SDGs irrespective of the order of paragraphs. We find differences between runs (for both models) and prompt versions (solely in the GPT3.5 case). While differences between prompt versions are expected – as questions differ and capture different aspects of the same problem – the order of appearance across paragraphs should not affect the classification of each text to important SDGs. As Gemini 1.0 maintains consistency across all runs and is not subject to randomness, we decided on Gemini 1.0 as the appropriate model for the study. We built a two-step prompt structure (Table 2) to capture relevant SDGs per paragraph and the tone of these connections. We avoid fine-tuning and domain-specific training as Gemini already includes knowledge about the SDGs.

| Table 2. Two-step prompt structure |
| --- |
| **First Prompt** |
| "Assign the following text to all relevant SDGs (strictly from the following list: 1) No poverty, 2) Zero Hunger, 3) Good health and well-being, 4) Quality education, 5) Gender equality, 6) Clean Water and Sanitation, 7) Affordable and clean energy, 8) Decent work and economic growth, 9) Industry, innovation and infrastructure, 10) Reduced inequalities, 11) Sustainable cities and communities, 12) Responsible consumption and production, 13) Climate action, 14) Life below water, 15) Life on land, 16) Peace, justice and strong institutions, 17) Partnerships for the goals). If a paragraph tackles non relevant issues with respect to any SDG, assign 0". |
| **Second prompt** |
| Use the following rules to interpret a paragraph: Consider climate adaptation as the adjustment in natural or human systems in response to actual or expected climatic stimuli or their effects, which moderates harm or exploits beneficial opportunities. Also consider, climate mitigation as an anthropogenic intervention to reduce the sources or enhance the sinks of greenhouse gas. Assign to each paragraph one and one only number between 0, 1 and 2. Assign 0 if if the paragraph explains or present an action or a set of actions which pose concrete risks to at least one between climate adaptation and mitigation; assign 1 if the paragraph is neutral with respect to climate adaptation and mitigation and does not express or discuss any concrete opportunity or risk for the country; assign 2 if the paragraph explains or present an action or a set of actions which pose concrete opportunities to at least one between climate adaptation and mitigation. |

The prompt-engineering process benefitted from a trial-error ping-pong exchange between the research team members. The final prompt was reached after eight attempts. We find different sentiments across countries, with respect to social, economic of environmental SDGs (Figure 3). As an automatic evaluation is not feasible because of the lack of benchmark datasets in this field, we engaged in a manual screening of both SDG allocations and sentiments. This process has its limits, and we acknowledge that evaluation protocols should be developed.

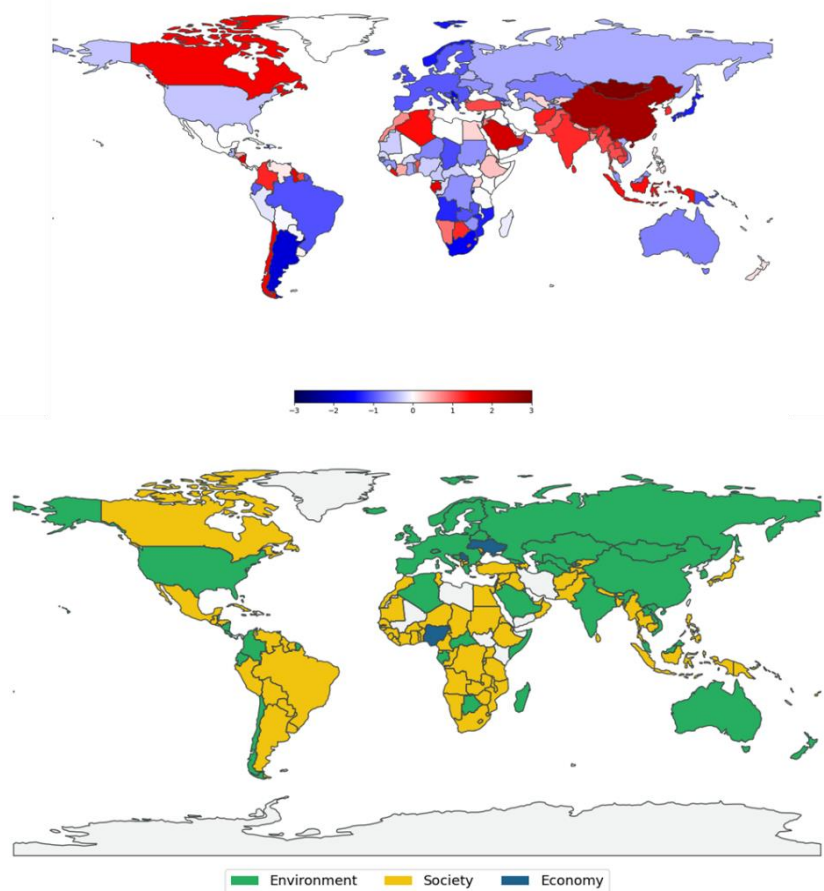

Figure 3. The geospatial distribution of sentiments and SDGs. Z-score of sentiment indicates the overall tone of countries in describing their actions with respect to climate adaptation and mitigation (upper part); SDGs are grouped according to their core interest (bottom-part). On the environment category: SDG6, SDG13, SDG14 and SDG15; in the society category: SDG1, SDG2, SDG3, SDG4, SDG5, SDG7, SDG11, SDG16; within the economy category: SDG8, SDG9, SDG10 and SDG12. We included SDG17 within the social category.

### 3.2. Detecting synergies and trade-offs among SDGs

In this second part of the study, we present the methodology behind a work aimed at detecting the type (positive or negative) and direction (outward or inward) of SDGs' interaction with one another using a large-scale corpus of policy and research documents. Our approach proposes full integration of human expertise and machine intelligence to guide action and inform policies around sustainable development worldwide. The term "sustainable development" appeared for the first time in the famous Brundtland report in 1987[37]. In there, "sustainable development" is defined as the ability of humanity to ensure that the needs of the present do not compromise "the ability of future generations to meet their own." Sustainable development is framed as a "goal" for everyone which the world shall reach by 2000. Since the publication of the Brundtland report, multilateral organisations (e.g., the United Nations), supranational institutions (e.g., the European Commission and the African Union), public and private actors have thrived towards realizing this promise using "sustainable development" as their guiding framework.

While being a concrete aspiration, "sustainable development" is also a multifaceted and complex concept. Since the publication of the Brundtland report, it lays on four critical dimensions: promoting economic growth, ensuring the respect and protection of the environment to avoid degradation, enabling everyone to fulfil her/his potential within communities, and supporting adequate institutions[37]. Policies (meant as the set of actions decision-takers put in place) supporting sustainable development may cover individual dimensions or them all. They can change overtime and – depending on specific contextual factors – they can target certain groups more than others. This high degree of heterogeneity is also a challenge to policy makers, which soon after the launch of the Brundtland report realized the world needed an operational guidance to fulfill the promises of sustainable development. In early 2000s, the international community presented the Millennium Development Goals (MDGs) which provided a roadmap to prosperous development by 2015 focusing on lower income countries. The MDGs

– signed by 189 countries – were centered on 8 goals, monitored through 18 targets and 48 indicators. Their progress enabled the world to half extreme poverty in fifteen years. However, their scope was not universal, but focused on low-income countries. For this reason, in 2015, the MDGs left space for the Sustainable Development Goals. Launched in Addis Ababa 2030 Agenda for Sustainable Development, the SDGs cover 17 goals, eight of which totally new with respect to their MDG ancestors, and they represent a global framework for the near-term future. Currently, the SDGs constitute the backbone of action towards a sustainable "common future": they shape the channels through which plans are designed and decisions are taken. As they describe pathways of change through their goals, targets and indicators, the SDGs are the most suitable tools to represent how different sustainability dimensions interact with one another. Their interlinkages (positive or negative) describe this complex landscape and – for this reason – they are the preferred lens through which the sustainable development agenda takes shape[38,39].

The study of synergies and trade-offs across sustainability domains has gained traction through the assessment of published evidence on SDG interlinkages. Given the heterogeneity in terms, focuses and interests around sustainable development, surveying the body of published (i.e., academic and non-peer-reviewed) knowledge is a challenge as no query will ever cover the entire universe. The literature has approached this problem by involving experts to pre-select studies based on their knowledge[5]. Over the past two decades, this trend has raised questions about the risk of oversimplifying reality[40], the need for credible, salient and legitimate information[41] and the representativeness of the knowledge systems these experts cover[42]. Calls to broaden inclusivity by using collaborative platforms, automated routines[43] and AI[44] have appeared, but the institutional processes are still digesting these routines and are far from implementing them. Instead, the dominant view lies on the creation of Expert Groups (e.g., the UNDESA Expert Groups Meetings for every SDG), which regularly meet across different sustainability dimensions or focused solely on specific areas. These groups shall not be dismantled as they contribute to those strong multilateral institutions which advance "sustainable development"[37], but they can make use of automated knowledge retrieval processes[2] so to focus on shifting from useful to usable insights[45].

To overcome the limits of potentially partial queries, we used a collaborative approach named Delphi Method. Originally developed in business studies, the Delphi Method aims at gathering knowledge from experts in the presence of diverse and potentially diverging opinions and expertise. We surveyed the authors of the manuscript covering engineering, computer science, mathematics, economics and sustainability sciences and we asked them to engage in an iterative process. The experts involved were asked to provide answers to a questionnaire without communicating to each other. In this specific case, the experts were asked to list all relevant documents according to their priorities. After the first round, the experts were presented with a partial consensus for them to review. The second stage led to a convergence towards a set of different policy documents. In the third and final stage, the experts collectively discussed the outcome refining whenever needed. The Delphi Method led to the identification of key document types as follows: i) peer-reviewed papers published from 2015 to identify academic backbones; ii) national reports on SDG progress to represent political actions; iii) civil society initiatives around the SDGs to test whether communities are advancing on selected issues and iv) news items to capture the pulse around sustainability in a given country.

As the heart of our research interest is the identification of synergies and trade-offs with clear directionality between different SDGs, we used a multi-stage prompt (Table 3).

| **Table 3. Multi stage prompt** |
|---|
| First Prompt: Identifying SDGs |
| ** Text ** : <br>      '{text}' <br><br>      ** Instructions **: <br><br>      Assign each index in this text to the main Sustainable Development Goal, as well as every other relevant but secondary SDG (strictly from the following list: 1) No poverty 2) Zero Hunger 3) Good health and well-being 4) Quality education 5) Gender equality 6) Clean Water and Sanitation 7) Affordable and clean energy 8) Decent work and economic growth 9) Industry, innovation and infrastructure 10) Reduced inequalities 11) Sustainable cities and communities 12) Responsible consumption and production 13) Climate action 14) Life below water 15) Life on land 16) Peace, justice and strong institutions 17) Partnerships for the goals). If a paragraph tackles non relevant issues with respect to any SDG, assign 0. <br><br>      ** Context **: <br><br>      Use the UN Sustainable Development Goals (SDGs) definitions in order to assign all the relevant SDGs to the text given above. Here are the definitions: {Read_description} |
| ** Output Format **: |
| Main SDG (pertinent number): Name of main SDG |
| Reason main SDG (pertinent number): clear justification as to how the SDG is the main SDG for the given text |
| |
| Secondary SDG (pertinent number): Name of SDG |
| Reason secondary SDG (pertinent number): clear justification as to how the SDG is pertinent but secondary to the given text |
| |
| Secondary SDG (pertinent number): Name of SDG |
| Reason secondary SDG (pertinent number): clear justification as to how the SDG is pertinent but secondary to the given text |
| |
| … and so on with as many pertinent but secondary SDGs for the given text. |
| """ |
| Second Prompt: Evaluating relationships. |
| ** Original Text **: |
| Text to analyze: "{text}" |
| |
| ** SDG Description **: {Read_description} |
| ** Instructions **: |
| |
| 1. **Evaluate Inter-relationships:** For each pair of SDGs identified within a text, determine whether they exhibit: |
| - **Synergy:** The SDGs positively reinforce or complement each other, contributing to a shared outcome. An example of a synergic relationship can be found in the following text "Climate change will increase the agricultural yield by 15%". |
| - **Trade-off:** The SDGs potentially conflict or compete with each other, where achieving one goal might negatively impact another. An example of a synergic relationship can be found in the following text "Climate change will increase poverty by 64% over the next 5 years". |
| - **Neutral:** The link between the SDG Pair is not clear in the context of the text. |
| |
| 2. **Identify directionality of inter-relationship:** For each synergy or trade-off identified, find explicit directionality of impact: |
| - Assign "outward" if the direction goes from the first identified SDG towards the second. An example of an outward trade-off can be found between SDG13 and SDG1 in the following text: "Climate change will increase poverty rates by 10% globally". |
| - Assign "inward" if the direction goes from the second identified SDG towards the first. An example of an inward trade-off can be found between SDG1 and SDG13 in the following text: "Climate change will increase poverty rates by 10% globally". |
| - Assign "both" if the direction goes both from the first to second and from the second to the first identified SDG. An example of a both trade-off can be found between SDG13 and SDG1 in the following text: "Climate change and poverty rates will increase by 10% globally". |
| |
| ** Output Format **: |
| |
| - SDG Pair: SDG SDG {main_sdg[0]} - {secondary_sdg} |
| - Relationship: [Synergy/Trade-off/Neutral] |
| - Directionality: [Inward/Outward/Both] |
| - Explanation: Using extracts from the text, reason how the SDG Pair constitutes a synergy or trade-off or neutral. Also, explain your Directionality assignment. |

In this table, the definitions of the SDGs' are given as part of the context windows, using Gemini's capacities. The definitions were downloaded from the United Nations website.

As in the first study, the prompt is the outcome of a trial-and-error procedure. In this specific case, we test the performance of each attempt on a baseline. We download the KnowSDGs database (https://knowsdgs.jrc.ec.europa.eu/interlinkages/advanced-search) created and maintained by the Joint Research Center (JRC) where experts collected documents and assigned each to up to two SDG targets indicating the type (synergy or trade-off). In absence of a dedicated labelled database, we deem the JRC's portal as the most appropriate as it includes documents as derived from diverse methods of study (e.g., data analysis, literature review, expert judgement, modeling, and semantic analysis). Preliminary results are promising as approximately 80 percent of the "predicted" (i.e., detected by the AI routine) SDGs match with the experts' assignation. We find indication of some imbalances across the SDGs, with SDG17 chronically underdetected and SDG1 spotted in more text excerpts than expected.

## 4. Discussing opportunities and limits

When used, machine learning and AI were proved useful at extracting relevant information and at producing new knowledge for selected sustainability dimensions: climate research[27,46,47], medicine[48], food security[49] and water systems[50] among others. Experts agree that AI can accelerate the achievement of the SDGs[2], but few published evidence has led to the discovery of non-obvious connections across SDGs. Among studies using NLP techniques, many have not engaged in deploying last generation LLMs. We move beyond the current state of the art and we present the process behind two studies tackling climate and sustainability policy. As stated by Schäfer and Hase[31], "reflexive yet integrative views" are needed to advance computational research, especially in social and climate sciences. Following this line of reasoning, we engage in constructive criticism on the strengths and weaknesses of the use of LLMs (*reflexivity*) and we offer future avenues to improve the field (*integrative*).

In the two parts of this study (Section 3.1 and 3.2), LLMs have introduced text-based approaches into nowcasting techniques to support decisions. By doing so, they widen the opportunities for alternative data collection in a *complementary* way. In the case of the NDCs, traditional content analysis approaches have tested the ambition efforts and the implementation of pledges to reach the Paris target. LLMs introduce a new angle: they are useful to instantly extract complex and non-homogeneous information supporting countries with limited data collection campaigns. LLMs further elicit non-obvious and hidden connections between sustainability domains. In the second part, our routine leads to the identification of many more connections than the baseline, where the experts were limited to two SDGs per paragraph. The computational power behind large-scale data analyses is such to move beyond the restrictions imposed by time-consuming studies and enables a more detailed identification of non-trivial relationships. Finally, the use of LLMs for climate and sustainability policy ensures that better and more detailed monitoring efforts are in place. Accountability and transparency are two essential ingredients of the trust-building process between countries and with civil society. By leveraging on documented coding routines and on replicable protocols, policy makers can check on progress and gaps in their development planning.

Some of the biggest challenges associated with the present framework lie in the data collection and labelling (Figure 4, Step 1). We identify three core issues, which could lead to a lack of interpretability of the outputs if not properly accounted for. First, peer-review works take time to publish and largely focus on identifying alignments across different policy agendas finding mainly synergies (especially within SDGs). This skewed distribution of the results is motivated by the genuine desire to support policy with proactive recipes for change rather than exposing bottlenecks. Furthermore, the literature on local evidence is scattered and hard to access: grey literature and ad-hoc reports are dominated by context-specific findings which may not always find space in international journals. As there is no global database for such documents yet, individual researchers must make choices which profoundly affect the outcome of their scientific enquiry. A large body of evidence can be found in projects' deliverables, financed activities and even press releases. These insights shall not be neglected as they capture the reality on the ground. Second, documents may often lack accessibility due to restrictive paywalls. This is especially relevant for disadvantaged institutions and contexts where large-scale agreements do not exist. Open-science practices are not the standard yet and the least developed countries suffer greatest limits. The same issues apply to corporate reports, which are offered under subscription and license agreements. Depending on data and text cleaning and retrieval routines, the outcome may be biased threatening external validity, meant as the extent to which results may be adequately generalized. Beyond access, restrictions to the use of LLMs on text corpora also happen to be in place for several publishers reducing the potentialities for full-scale analyses. Limits to interpretability are also posed by the queries under use. This is relevant for linguistic purposes and for issues which change overtime. An example is climate policy research, which has evolved over the past four decades. Policies were centered on "global warming" and "greenhouse gas effect" back in early 2000s. As the science has evolved, so has the way we refer to these phenomena, favouring "climate change", "climate collapse" and "climate crisis". A comprehensive query aiming at capturing the full spectrum of a given sustainability domain may need to adopt all three modifications.

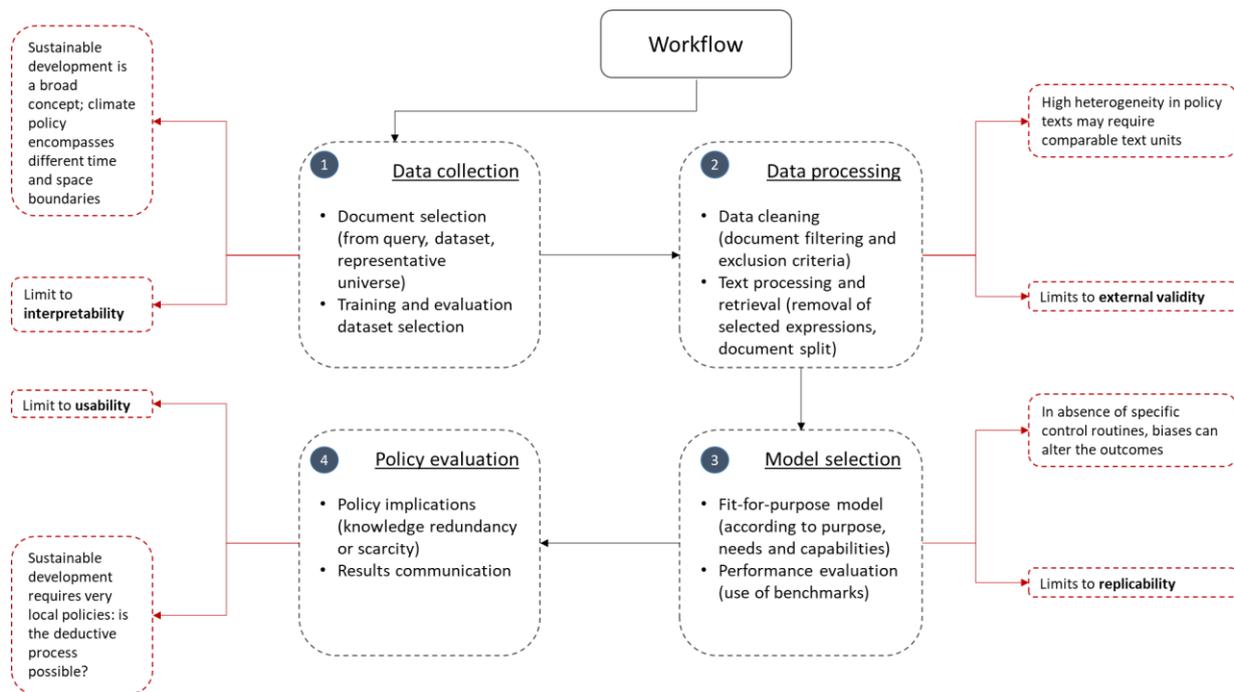

Figure 4. The application of LLMs to sustainability and climate policy documents in a four-step workflow. Each step presents a specific threat (or limit).

Beyond data collection, data processing (Figure 4, Step 2) poses the second challenge in applying automated routines (and particularly LLMs) to sustainability and climate policy documents. In the presence of big literature and big data queries, researchers may need to classify documents (according to time and space dimensions) removing those documents which do not seem fit and may skew distributions of topics, issues or themes of interest. This data collection procedure is different from text cleaning and text retrieval processes. As highlighted in Section 1, the desire to capture the meaning of complex texts grounds the deployment of content and discourse analysis and evolved into computational linguistics. Meaning can change depending on the unit under consideration: the full document conveys a message which may be different from the one presented in a single sentence, paragraph or chapter. In the process of analysis of SDG interlinkages, these differences become crucial and require strong reflections about the optimal context-window length. Many of the pre-GPT-era LLMs were trained at sentence level, implicitly forcing researchers to engage in sentence-level embeddings. It is with the post-GPT launch that modifications of the unit of analysis became possible. The release of larger models came with an increased context window enabling in same cases (*i.e.*, Gemini 1.5 and foreseen GPT5) the scan of a full length text. The choice of the most appropriate context window strongly impacts the identification of synergies and trade-offs across the SDGs. Even when the SDGs are correctly identified, their frequency changes, leading to unclear implications for policy. Text cleaning from special characters, recurring expressions, numbers and selected linguistic forms constitutes an important step in NLP applications (mostly frequency-based statistical approaches). LLMs (i.e., from GPT3 onwards) are less affected by irrelevant expressions as the self-attention mechanism enables modeling the influence of every word on the others in parallel, through the computation of an attention score what every word means in context aided by a wider and more elaborate context-window. Still, data cleaning remains a good practice to enable full-scale reproducibility and involves also taking decisions about inclusion and exclusion rules of figures and tables headings and captions, titles and subtitles and focus boxes.

Both interpretability and external validity are limits encountered and documented in the use of LLMs in clinical and medical research[51,52]. In the sustainability and climate policy domains, we acknowledge two additional risks identifiable in the model selection and output evaluation stages (Figure 4, Step 3). The choice of the most suitable and fit-for-purpose model is conditioned by market forces and four other factors. There are plenty of LLMs available and released at impressive speed[53]. Literature has classified LLMs according to the their neural architecture[53], pre-training method[54], accessibility, general purpose and domain-specific type, multilingual and mono, bilingual abilities[55], among others. The choice of the model used to solve the specific research question depends on several factors:
- Technical – context window, prompting technique and available computing resources.

- Economic - open-source and open-access models are preferred alternatives, especially in budget-constrained contexts.
- Policy - interpretable outputs and full-scale replicability are core needs and the use of "black boxes" needs to be avoided[15].
- Purpose: the objective is highly context- or domain-specific and requires fine tuning.

As models are published and used, evaluations and applications flourish. Platforms such as HuggingFace OpenLLM Leaderboard (https://huggingface.co/spaces/open-llm-leaderboard/open_llm_leaderboard) provide benchmark evaluations to assess model capabilities in a standardized way. When certain models exhibit worse performance than others, they may be dismissed. Finally, certain models may be withdrawn because of issues encountered in their initial deployments or simply retired and not made available anymore (*e.g.* Gemini 1.0). These market-driven factors, together with the technical, economic, policy and purpose motives, put at risk the reproducibility of results and ultimately the legacy of the derived policy implications. Prompting techniques in sustainability and climate policy analyses overcome these challenges in two ways. Prompts support replicability and they can be designed to respond to specific purposes. Even in absence of a specific model, the process and workflow is documented. As prompts are typically in use for general-purpose LLMs - trained on billion parameter scale large text corpora – their use surveys the world wide available knowledge. To reach their objective, prompts must be specific, calibrated on the research question and elaborated enough to cover the objective. Prompts in sustainability and climate policy may need refinement over several iterations and they may involve multiple experts in domain-specific knowledge compartments.

The final stage of the workflow relates to the translation from useful to usable science, from research into policy and action (Figure 4, Step 4). This step is probably the most complex as it involves different skillsets. Communication of the results is the first milestone: as there is a gap between what researchers understand as useful and what users recognize as such[56], the presentation of the insights gathered from the research shall be co-produced with relevant stakeholders. When appropriate, uncertainties in the model outputs and human-made decisions which potentially lead to risks and threats shall be made explicit. Limits to usability of the outputs are also imposed by the scale of the outputs. Global or regional findings can provide information on trends and patterns of change. However, local decision-makers may require higher granularity to design policies which catch the specificities of the context. The effort to link global, regional and local settings must come from the researchers themselves and represents an important portion of the task.

## 5. Implications and concluding remarks

We find that there is a need for urgent research and development in three core areas. First, the use of LLMs in sustainability and climate policy cannot be separated from the development of transparent output-evaluation methods. In tasks such as text classification, sentence similarity and zero-shot classifications the assessment of LLM results is an important step to assess whether some dimensions or event topics are extensively or poorly tackled by certain policy interventions. The abundance of specific areas in policy documents can – for example – lead to stronger funding allocations as policy makers can perceive them as urgent and prioritize those issues. On the other end, topics which appear less urgent can remain critically left out, exacerbating inequalities and unjust dynamics. Second, LLMs extract meaningful insights considering the semantic and syntactic processes which give birth and shape to a text. LLMs are also trained on large-scale corpora. However, the assumption that LLMs grasp the meaning of policy documents is only partially correct. Being the outcome of compromise, negotiations and often long iterative processes, policy texts are expressions of what cannot be made explicit: intent and hidden motives. In the climate domain this anecdotal evidence has been widely documented[57] and raises questions about the actual comparability between documents produced in different moments or signed and put forward by different countries, departments or agents[58]. Third, the use of LLMs requires significant human interventions and decisions which may often be overlooked. While the use of keyword-based methods has been superseded and conceived as strongly biased, the deployment of prompts may suffer from the very same threats. In one of our applications, we tested how subtle but meaningful differences play a role in deciding over the use of one LLM over another. These details are far from being insignificant in policy processes: they make their mark in interpretation of the results. Equally important, we noted that while the launch of LLMs may require limited specialized knowledge, the sense-making of the insights they produce calls for highly expert-based skills. Therefore, we advocate for a new paradigm where human-aware machine-enabled insights become an integral part of the policy processes.

The discussion around limits and opportunities of LLMs in climate and sustainability policy reveals that more interdisciplinary research is needed. As LLMs are becoming more powerful, we can confidently expect that accuracy increases too. However, technical progress shall be complemented with more structured data collection campaign to inform AI-based routines. Also large-scale efforts in assessing the most suitable models and protocols can support the identification of policy-relevant routines. The ClimateChange AI community and the AI for Good Summit are experiences which teach how collaboration across diverse scientific domains can drive the change. Finally, the beneficiary itself (policy) shall invest more in these tools. Society calls for quick, scalable and coordinated transformations where private and public actors come together supporting one shared goal. Currently, AI is one of the many tools available to inform and shape evidence making processes. However, to move from useful to usable insights, uptake is essential. Starting with AI literacy, policy makers can enhance their decisions by using nowcasted large-scale results. If complemented by their expertise, experience and political agenda AI methods can enhance science and accelerate change.

**Competing interests**
The authors declare no competing interests, financial or other, exist.

**Funding statement**
RV, FFN and FL acknowledge the funding provided by Digital Futures, in their Demonstrator-project program. FL acknowledges funding from the European Union's Horizon Europe research and innovation programme under the Marie Skłodowska-Curie grant agreement No. 101150729. SH acknowledge the grant PID2021-128676OB-I00 funded by MCIN/AEI/10.13039/ 501100011033 and by "ERDF A Way of Making Europe", by the European Union. JAC acknowledges funding from the program Ayuda a Primeros Proyectos de Investigación (PAID-06-23) from the Vice-rectorate for Research from Universitat Politècnica de València (UPV).

**Acknowledgements**
The team acknowledges the support and constructive feedback of Oscar Garibo Orts and Fermin Mallor Franco.